\documentclass[reprent,notitlepage,jcp]{revtex4-1}

\usepackage{dcolumn}
\usepackage{bm}
\usepackage{geometry}
\usepackage{graphicx} 

\usepackage{float}
\usepackage{multirow}
\usepackage{booktabs} 
\usepackage{array} 
\usepackage{paralist} 
\usepackage{verbatim} 
\usepackage{subfig} 
\usepackage{rotating}
\usepackage{rotating}
\usepackage{tikz}	
\usetikzlibrary{shapes,snakes}
\usetikzlibrary{backgrounds,fit,decorations.pathreplacing}  

\usepackage{fancyhdr} 
\usepackage{threeparttablex}
\usepackage{amsmath, amsthm}
\usepackage{afterpage}
\usepackage[nottoc,notlof,notlot]{tocbibind} 
\usepackage[titles,subfigure]{tocloft} 

\graphicspath{{./figures/}}
\DeclareGraphicsExtensions{.eps}

\usepackage{caption}
\begin{document}

\title{ Quantum Confinement and  Negative Heat Capacity}

\author{Pablo Serra${}^{1,2}$, Marcelo Carignano${}^2$, Fahhad Alharbi${}^{2,3}$, and Sabre Kais${}^{2,4}$}\email{kais@purdue.edu}

\affiliation{${}^1$Facultad de Matem\'atica, Astronom\'{\i}a y F\'{\i}sica, Universidad Nacional de C\'ordoba and IFEG-CONICET, Ciudad Universitaria, X5016LAE C\'ordoba, Argentina}
\affiliation{${}^2$Qatar Environment and Energy Research Institute, Qatar Foundation, Doha, Qatar}
\affiliation{${}^3$King Abdulaziz City for Science and Technology, Riyadh, Saudi Arabia}
\affiliation{${}^4$Department of Chemistry,  Physics and Birck Nanotechnology Center, Purdue University, West Lafayette, IN 47907 US}


\begin{abstract}
Thermodynamics dictates that the specific heat of a system is strictly non-negative. However, in finite classical systems there are well known theoretical and experimental cases where this rule is violated, in particular finite atomic clusters. Here,  we show for the first time that negative heat capacity can also occur in finite quantum systems. The physical scenario on which this effect might be experimentally observed is discussed. Observing such an effect might lead to the design of new light harvesting nano devices, in particular a solar nano refrigerator. 

\end{abstract}

\maketitle


Thermodynamics dictates that the specific heat of a system is strictly non-negative, implying that the addition (subtraction) of energy cannot result in a decrease (increase) of the system's temperature \cite{huang-statmech}. Nevertheless there are well known cases where this rule is apparently violated \cite{Behringer:2005aa,Komatsu:2009aa}. For example, this phenomenom occurs at the nanoscale and it was eloquently showed by Schmidt et al. who, in an elegant series of experiments, determined a negative heat capacity of a Na cluster with 147 atoms for temperatures neighboring the melting temperature of the cluster \cite{Schmidt:1997aa,Schmidt:1998aa,Schmidt:2001aa}. 
Also at the astronomical scale negative heat capacities have been known for years \cite{LYNDENBE.D:1968aa}, where it is observed that stars and star clusters increase their temperature as they age while loosing energy by radiation \cite{Lynden-Bell:1999aa}.  Therefore, invoking the thermodynamic limit is not sufficient to guarantee the equivalence of Canonical and Microcanonical ensembles.
The key to theoretically reconcile these results with the thermodynamics was addressed by Thirring and coworkers: A system may display negative heat capacity, even in the thermodynamics limit, provided that it is not ergodic \cite{THIRRING:1970aa,HERTEL:1971aa,Thirring:2003aa}.
The ambiguity of negative heat capacity concept has been extensively examined by the work of RS Berry and coworkers \cite{Berry:2009a,Berry:2006a,Jortner:2006a,Berry:2010a,Berry:2004a,Berry:2004b} . In this letter we investigate whether this phenomenon could also be observed in the quantum domain, in particular, for small systems. An important motivation for the understanding and control of this phenomenon would be the implementation of refrigeration strategies at the nano scale, for example for light harvesting nano devices that would benefit from an inverted thermal response.

Following previous ideas \cite{Rao:2008aa,Carignano:2010aa} on a minimal model having negative specific heat for classical systems we study the effect of the delocalization of the wave function on the average kinetic energy of the system, and also on several definitions of temperature corresponding to the Canonical and Microcanonical statistics. 
Let us consider a 1d potential well trapped between impenetrable walls:
\begin{equation}
\label{pot}
V(x)\,=\,\left\{ \begin{array}{lll} -U_0 & \mbox{if}&  |x| < a \\
0 & \mbox{if } & a\leq |x| < L \\
\infty
&\mbox{if }& |x| \leq L\\
 \end{array}  \right. \,,
\end{equation}
This potential represents a simple example that suffices to show how a negative heat capacity emerges. The solution of the Schr\"odinger equation for one particle in $V(x)$ given by Eq. (\ref{pot}) can be found in ref \cite{flugge}. The  solution for a system of $N$ noninteracting electrons can be constructed from the one particle solution. The Coulombic correlation effects for $N=2$ were tested to be very small and therefore neglected in the calculations for higher $N$. The values for the parameters $U_0$ and $a$ for the external potential $V(x)$ were selected to approximately represent quantum dots binding energy and size and are specified in the corresponding figures. 

Within the Canonical formalism there is no ambiguities to establish the relation between the energy of the different eigenstates and the temperature of the system, and a continuos temperature can be assigned to a Canonical ensemble. However, in the Microcanonical ensemble the situation is different and to the best of our knowledge there is no accepted 'recipe' to calculate the temperature corresponding to a particular eigenstate. Here we will define the entropy of a Microcanonical system as $S(E)= k_B \ln \Omega(E)$, where $\Omega(E)$ is the number of states with $E_n \leq E$ and $k_B$ the Boltzmann constant. $S(E)$ is a piecewise constant function that eventually converges, as it will be shown below, to a continuous convex function of the energy as the number of particles goes to infinity. Straightforward application of the thermodynamic definition of temperature, $1/T^{(\mu )}=\partial S/\partial E$, to our definition for $S(E)$ does not produce a finite temperature. However, the general structure of $S(E)$ from different examples suggests that a slope ({\em i.e.} a temperature) can be derived from this curve. Here we propose to calculate the Microcanonical temperature using finite difference between adjacent states:
\begin{equation}
T^{(\mu )}=\frac{\Delta E_n}{\Delta S_n}.
\label{fdt}
\end{equation}
This definition has the disadvantage that it is not unique, since the finite difference could be taken in many different ways, being the forward, backward and centered differences the more usual choices in several fields. Moreover, this definition proves to be useful for $N=1$ but it becomes noisy for larger $N$. Note also that from Eq. (\ref{fdt}) the temperature results a quantized property of the system, as it can be calculated only for the eigenvalues of the system. A second possibility for the definition of the temperature emerges from the kinetic energy operator $K$, which can be combined with the equipartition theorem to produce
\begin{equation}
T_n^{(K)}=\frac{2}{d} \frac{\langle n | K | n \rangle}{N} = \frac{2}{d} \frac{\langle K \rangle_n}{N}
\end{equation}
where $| n \rangle$ represents a pure state of a system of $N$ particles. 

\begin{figure}[!t]
\begin{center}
\includegraphics*[width=0.8\textwidth]{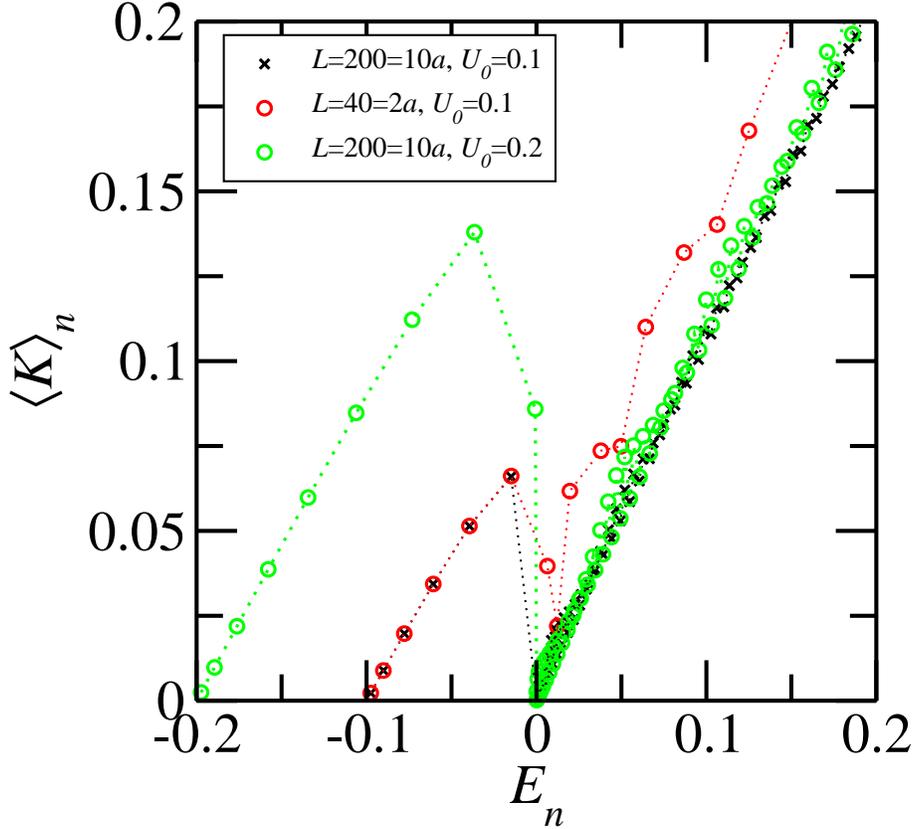} 
\end{center}
\caption{Expectation value of the kinetic energy as a function of the energy for different model parameters.  A drop in $\langle K \rangle_n$ occurs as the particle is able to explore regions beyond the central well.
}
\label{K-E}
\end{figure}

\begin{figure}[!t]
\begin{center}
\includegraphics*[width=0.8\textwidth]{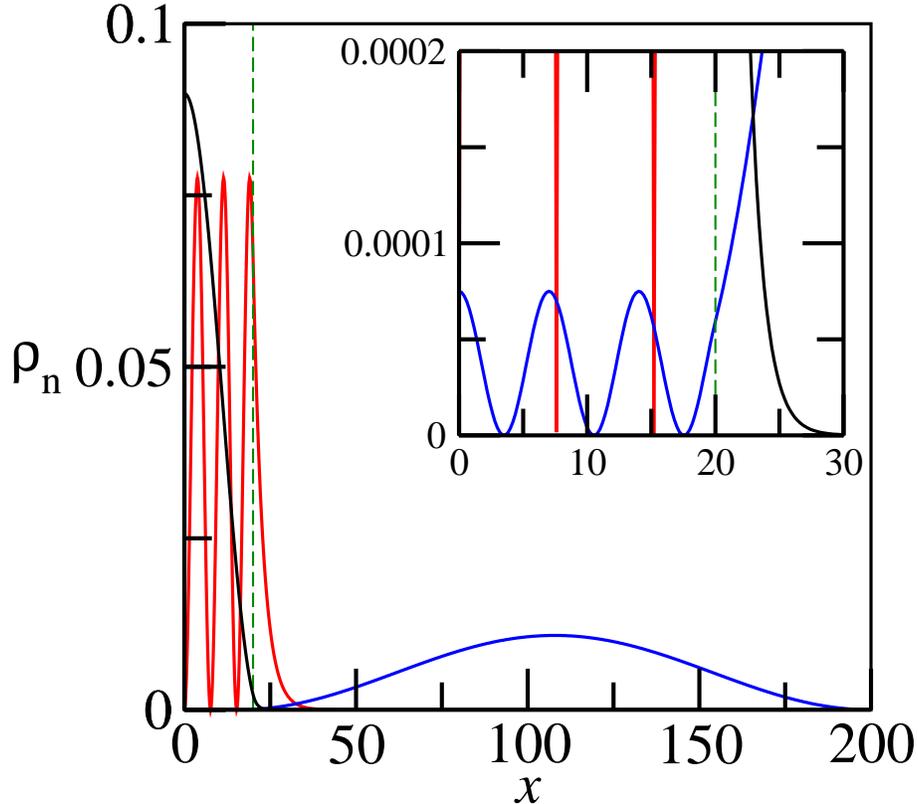} 
\end{center}
\caption{Probability density for $L=200=10 a$, $U_0=0.1$ for the ground state (black, $E_0=-0.0975$), highest negative (red, $E_5=-0.0149$) an lowest positive (blue, $E_6=0.000145$). 
}
\label{rho}
\end{figure}

In Figure \ref{K-E} we show $\langle K \rangle_n$ vs $E_n$ for one particle in the model potential of Eq. (\ref{pot}) for three different parameter sets. The parameters displayed in the figure are in atomic units and $a=1$, and the examples shown correspond to a central confinement region with a diameter of 0.2 nm and 1 nm and a well depth of 2.7 eV and 5.4 eV. This values are representative of multi-layered semiconductor quantum dots \cite{ferron:2012aa,ferron:2013aa}.
All three curves shows the same qualitative feature: as the energy approaches the threshold value to escape from the central well, $\langle K \rangle_n$ decreases implying a negative heat capacity. The magnitude of this effect depends on the depth of the well $U_0$ that for a fixed $a$ controls the number of bound states and therefore defines the lower energy branch of Figure \ref{K-E}, and the relative delocalization $L/a$ that controls the magnitude of the reduction in kinetic energy as the particle exits the well. This dependence is very similar to what occurs in the classical case.
The reason for this becomes clear by looking at the probability density for a particular case. 
For model potential of the general shape as the one described by Eq (\ref{pot}), the origin of the temperature decrease in the classical problem is the availability of a large space accessible upon the increment of a small amount of energy. In the quantum case the same general picture is valid, with a minor adaptation since a  quantum particle is able to penetrate the regions of negative energy due to exponential tail of the wave function. Namely, for a quantum particle the delocalization starts to take place for those eigenstates with $E_n$ slightly smaller than zero. In Figure \ref{rho} we show the probability density for the ground state together with the highest state just below the edge of well expansion and lowest state just above that. The probability density of the ground state and the next three states (not shown) are {\em almost} fully localized in the region $|x| \lesssim a$ and therefore the kinetic energy follows the expected thermodynamic behavior, namely, it linearly increases with the system energy. The highest bounded state (negative energy) shows a small probability to be found for $|x| > a$ and since this region contributes with a much smaller kinetic energy the expectation value $\langle K \rangle_n$ displays a bend departing from the linear relation. The first state with positive energy can explore with no penalty (meaning the wave function is oscillatory, not exponentially decaying) the full available space. This possibility to leave the central well is reflected in a much lower $\langle K \rangle_n$, which in turn  imply a negative heat capacity if we directly associate kinetic energy with temperature. Further increase of the energy once the threshold value has been crossed results again in the expected monotonic increase of the kinetic energy. From Figure \ref{K-E} it can also be observed the emergence of a noise, as an odd-even effect for  $E_n > 0$, in particular for the case $L=200=10a$ with $U_0=0.2$.

\begin{figure}[!t]
\begin{center}
\includegraphics*[width=0.8\textwidth]{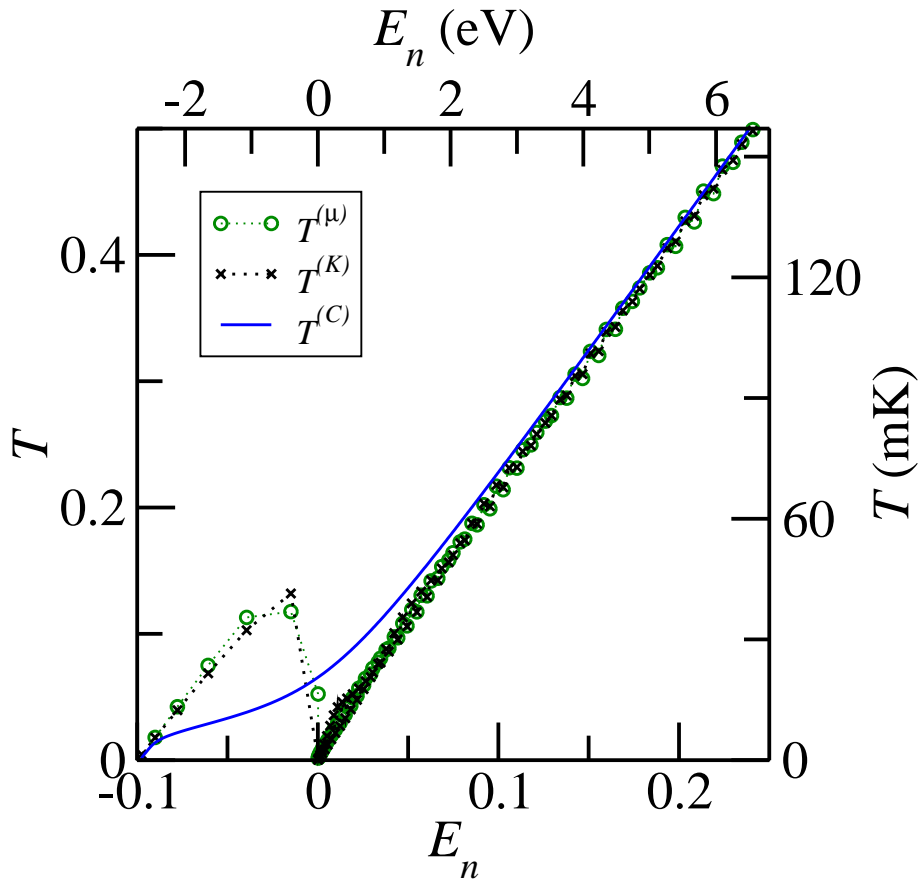} 
\end{center}
\caption{Temperature vs. energy for the parameters corresponding to Figure \ref{rho}. All temperature definitions are equivalent for large energy values.
The upper and right axis units are calculated assuming $U_0=2.7$ eV, $a=1$ nm, $L=10$ nm.
}
\label{T-E-onebody}
\end{figure}

We now compare the predictions of the Microcanonical and Canonical ensembles, $T^{(\mu)}$ and $T^{(C)}$, as well as the temperature defined from the kinetic energy operator $T^{(K)}$. In Figure 3 we show these temperatures corresponding to the case displayed in Figure 2. First we note the remarkable similarity between $T^{(K)}$ and $T^{(\mu)}$. This result is somehow surprising because the former is obtained from the kinetic energies of the pure states, while the later is a result of the structure of the spectrum from the ground state up to the level being considered. It is important to mention that in the calculation of $T^{(\mu)}$ we have used the centered difference formula. If the forward or backward finite difference derivatives are employed the qualitative results are the same, but those curves display a characteristic odd-even effect that introduce an unnecessary complication.
The Canonical temperature $T^{(C)}$ shows the expected thermodynamic behavior as it increases monotonically with the energy of the system, smoothly bridging the low and high temperature regimes where it matches the other two temperature definitions. 
Figure 3 clearly shows that a quantum negative heat capacity emerges from the model potential much in the same way that it does for classical systems. The sudden availability of space, accessible upon a small increment of the total energy, leads to the decrease of the average temperature of the system.

\begin{figure}[!b]
\begin{center}
\includegraphics*[width=0.8\textwidth]{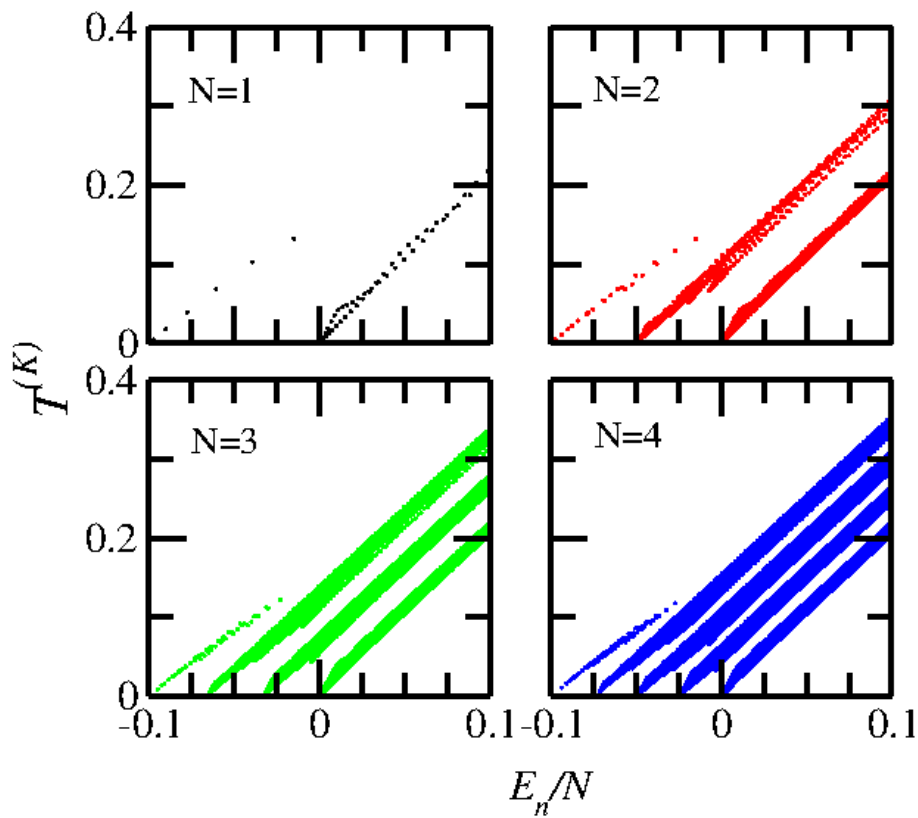} 
\end{center}
\caption{$T_K$ as a function of the energy per particle for systems of 1 to 4.
}
\label{tkmany}
\end{figure}

\begin{figure}[!t]
\begin{center}
\includegraphics*[width=0.8\textwidth]{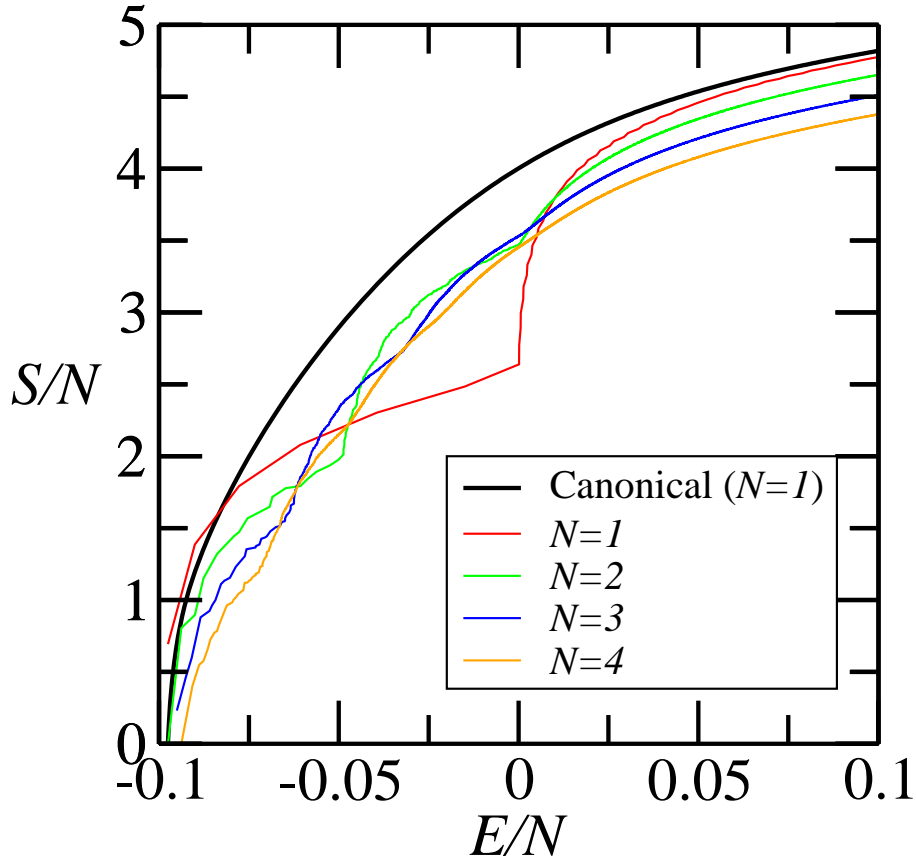} 
\end{center}
\caption{Entropy per particle as a function of the energy per particle for systems with a few particles. Increasing $N$ reduce, and eventually eliminates, the concavity region while approaching the Canonical result.
}
\label{entropy}
\end{figure}

The question that arises is how this results are affected by increasing the number of particles in the system. The definitions of $T^{(K)}$ and $T^{(\mu)}$ are independent of the system size and therefore may be directly applied to the problem of several particles. In Figure \ref{tkmany} we show $T^{(K)}$ as a function of energy for $N=1, 2, 3$ and $4$. For $N = 1$, $T^{(K)}$ shows only one jump as the particle leaves the central well. However for $N>1$ the kinetic energy shows successive drops as the total energy allows each of the particles to explore the region with $|x| > a$. Then, for $N=2$ there appear two branches in the $T^{(K)}$ vs $E$ curve, three branches for $N=3$, and so on. This behavior eliminates $T^{(K)}$ as useful generic definition for the temperature of a system. The calculation of the  Microcanonical temperature $T^{(\mu)}$, although possible, does not provide a clear vision of the behavior of the system because of the noise introduced by the finite difference method. Nevertheless, the Microcanonical entropy calculated by simply counting the number of states allows a different and clearer interpretation. In Figure \ref{entropy} we show our results for $S$ vs. $E$ for systems of up to $N=4$. The figure also includes the Canonical result, which is independent of the number of particles in the non-interacting case. Note that Figure \ref{entropy} shows the values of quantized entropy connected with straight lines, and as the number of particles increases the noise of the curve is reduced although not enough to result in a smooth curve upon numerical differentiation. For $N=1$ the entropy shows a concave kink at $E \simeq 0$ that leads to a temperature drop. As $N$ increases, this kink splits as many times as particles in the system. Each one of these $N$ kinks is weaker than the $N-1$ kinks of a smaller system. Therefore, it is clear that this kink splitting mechanism will eventually completely remove all the concavities in the $S$ vs. $E$ curve as $N$ becomes very large. As a consequence, in the thermodynamic limit, we expect a full equivalence of the Microcanonical and Canonical descriptions. In fact, the Canonical entropy also displayed in Figure \ref{entropy} is essentially parallel to the Microcanonical entropy for $N=4$. The fact the the two ensembles are equivalent on the thermodynamic limit is expected since the simple model of Eq. (\ref{pot}) does not contains energy barriers that could lead to  non-ergodicity, another source of difference between the ensembles.

The physical scenario on which this effect could be experimentally observed in the quantum domain is with quantum confined semiconductor devices, core/shell nanoparticles or even graphene layers \cite{Aoki:2007aa,Singh:2011aa} or nanoribbons \cite{Han:2007aa}.
The key ingredient is to have spacial confinement connected with a much wider region, i.e. a {\em localized} and a {\em delocalized} regions, where the access to the second becomes possible upon a small increment of the system's energy. 
Spatial delocalization can be in principle constructed in multiple quantum wells and wires and core/shell nanoparticles. By proper doping and design, the highly localized states can be occupied and electrons can be excited to the delocalized states by low-energy photons.
However, the challenge is how to measure the delocalization effect on the temperature. Carrier temperature has been estimated indirectly and qualitatively using optoelectronic measurements and micro-Raman spectroscopy \cite{IKOMA:1989aa,fahhad1,Xi:2005aa,TODOROKI:1985aa}. The success of an experimental verification of these predictions might lead to the design of new light harvesting nano devices, and in particular a solar nano refrigerator.

\section*{Acknowledgements}

The authors would like to acknowledge Prof. Timothy S. Fisher from Purdue University for discussions and his continuing efforts to harness this effect in an experimental system.

\end{document}